%% file: EAGER.tex
\documentclass[sigconf,authorversion]{acmart}

\copyrightyear{2024}
\acmYear{2024}
\setcopyright{acmlicensed}\acmConference[KDD '24]{Proceedings of the 30th ACM SIGKDD Conference on Knowledge Discovery and Data Mining}{August 25--29, 2024}{Barcelona, Spain}
\acmBooktitle{Proceedings of the 30th ACM SIGKDD Conference on Knowledge Discovery and Data Mining (KDD '24), August 25--29, 2024, Barcelona, Spain}
\acmDOI{10.1145/3637528.3671775}
\acmISBN{979-8-4007-0490-1/24/08}

\usepackage{subcaption}
\usepackage{multirow}
\usepackage{enumitem}
\usepackage{graphicx}
\usepackage{balance}
\usepackage{amsmath}
\usepackage{booktabs}
\usepackage{amssymb}
\usepackage[capitalize]{cleveref}
\usepackage{pifont}

\usepackage{colortbl}  %彩色表格需要加载的宏包
\usepackage{xcolor}
\usepackage{array}   %对表列和表格线的设置需要用到array宏包
\usepackage{marvosym} %Letter
\newcommand{\vpara}[1]{\vspace{0.05in}\noindent \textbf{#1 }}

\newcommand{\ie}{\textit{i}.\textit{e}.}

\newcommand{\xjh}[1]{\iffalse #1 \fi{\color{blue} \textbf{[XJH]}}}
\newcommand{\wy}[1]{\iffalse #1 \fi{\color{orange} \textbf{[WY]}}}

\definecolor{linecolor}{gray}{.91} % soft gray
\definecolor{linecolor2}{gray}{.95} % soft gray
\definecolor{linecolor1}{gray}{.97} % soft gray

\begin{document}

\title{EAGER: Two-Stream Generative Recommender with Behavior-Semantic Collaboration}

\author{Ye Wang}
\authornote{Equal contribution.}
\affiliation{%
  \institution{Zhejiang University}
  \city{Hangzhou}
  \country{China}
}
\email{yewzz@zju.edu.cn}

\author{Jiahao Xun}
\authornotemark[1]
\affiliation{%
  \institution{Zhejiang University}
  \city{Hangzhou}
  \country{China}
}
\email{jhxun@zju.edu.cn}

\author{Minjie Hong}
\affiliation{%
  \institution{Zhejiang University}
  \city{Hangzhou}
  \country{China}
}
\email{hongminjie@zju.edu.cn}

\author{Jieming Zhu}
\authornote{Corresponding authors.}
\affiliation{%
  \institution{Huawei Noah's Ark Lab}
    \city{Shenzhen}
  \country{China}
}
\email{jiemingzhu@ieee.org}
% Tao Jin, Wang Lin, Haoyuan Li, Linjun Li, Yan Xia, Zhou Zhao, Zhenhua Dong

\author{Tao Jin}
\affiliation{%
 \institution{Zhejiang University}
   \city{Hangzhou}
  \country{China}
 }
\email{jint_zju@zju.edu.cn}

\author{Wang Lin}
\affiliation{%
  \institution{Zhejiang University}
    \city{Hangzhou}
  \country{China}
}
\email{linwanglw@zju.edu.cn}

\author{Haoyuan Li}
\affiliation{%
  \institution{Zhejiang University}
    \city{Hangzhou}
  \country{China}
}
\email{lihaoyuan@zju.edu.cn}

\author{Linjun Li}
\affiliation{%
  \institution{Zhejiang University}
    \city{Hangzhou}
  \country{China}
}
\email{lilinjun21@zju.edu.cn}

\author{Yan Xia}
\affiliation{%
  \institution{Zhejiang University}
    \city{Hangzhou}
  \country{China}
}
\email{xiayan.zju@gmail.com}

\author{Zhou Zhao}
\authornotemark[2]
\affiliation{%
  \institution{Zhejiang University}
    \city{Hangzhou}
  \country{China}
}
\email{zhaozhou@zju.edu.cn}

\author{Zhenhua Dong}
\affiliation{%
  \institution{Huawei Noah's Ark Lab}
    \city{Shenzhen}
  \country{China}
}
\email{dongzhenhua@huawei.com}

% \thanks{\textsuperscript{\Letter} Corresponding authors.}

% \renewcommand{\shortauthors}{Wang and Xun et al.}
\renewcommand{\shortauthors}{Ye Wang et al.}

\begin{abstract}
Generative retrieval has recently emerged as a promising approach to sequential recommendation, framing candidate item retrieval as an autoregressive sequence generation problem. However, existing generative methods typically focus solely on either behavioral or semantic aspects of item information, neglecting their complementary nature and thus resulting in limited effectiveness. To address this limitation, we introduce EAGER, a novel generative recommendation framework that seamlessly integrates both behavioral and semantic information. Specifically, we identify three key challenges in combining these two types of information: a unified generative architecture capable of handling two feature types, ensuring sufficient and independent learning for each type, and fostering subtle interactions that enhance collaborative information utilization. To achieve these goals, we propose (1) a two-stream generation architecture leveraging a shared encoder and two separate decoders to decode behavior tokens and semantic tokens with a confidence-based ranking strategy; (2) a global contrastive task with summary tokens to achieve discriminative decoding for each type of information; and (3) a semantic-guided transfer task designed to implicitly promote cross-interactions through reconstruction and estimation objectives. We validate the effectiveness of EAGER on four public benchmarks, demonstrating its superior performance compared to existing methods. Our source code will be publicly available on PapersWithCode.com.
\end{abstract}



\maketitle

\input{part/wy} % 绪论
\input{part/xjh}

% \newpage

\balance
\bibliographystyle{ACM-Reference-Format}
\bibliography{EAGER}

\end{document}

%% file: part/wy.tex
\section{introduction}

% \begin{figure}[h]
%   \centering 
%   \includegraphics[width=0.5\linewidth]{figure/attn_diff.png}
%   \caption{{\bf Attention distance visualization of two recommendation paradigms: ID- and modality-based features.} }
%   \label{fig:attn_diff}
% \end{figure}

Recommender systems are widely adopted solutions for managing information overload, designed to identify items of interest to users from a large item corpus. Modern recommender systems typically integrate representation learning and search index construction to refine the matching process. Initially, users and items are encoded into latent representations within a shared latent space using models like two-tower architectures~\cite{MNS,CBNS} and sequential recommendation models~\cite{kang2018self,sun2019bert4rec}. Subsequently, to efficiently retrieve top-k items for users, approximate nearest neighbor (ANN) search indexes are constructed using tools such as Faiss~\cite{johnson2019billion} and SCANN~\cite{guo2020accelerating}. Despite notable progress, the separate phases of representation learning and index construction often operate independently, presenting challenges for achieving end-to-end optimization and consequently limiting the overall effectiveness of recommender systems~\cite{tay2022transformer}.

\begin{figure}[t]
  \centering 
  \includegraphics[width=.9\linewidth]{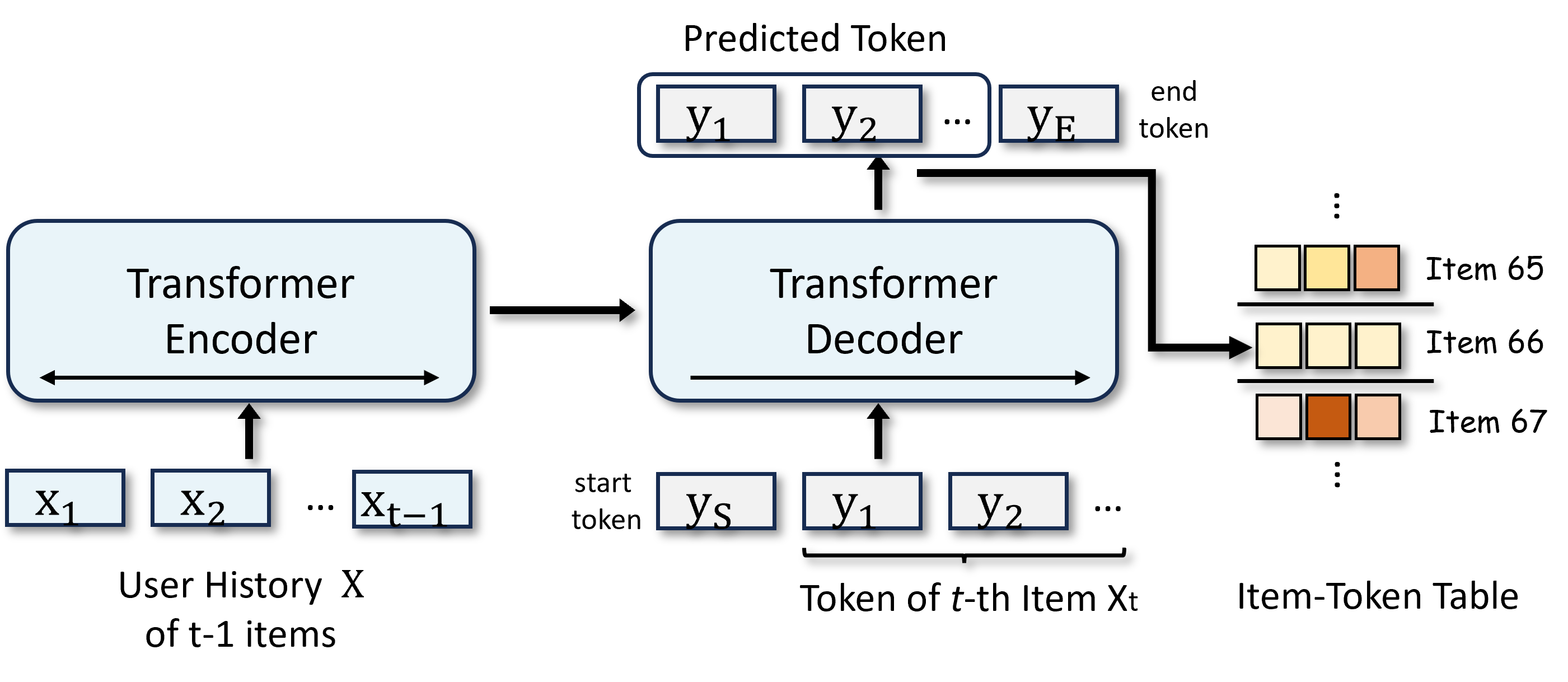}
  \vspace{-2ex}
  \caption{{\bf A framework of generative recommendation.} }
  \label{fig:intro}
  % \vspace{-.5cm}
\end{figure}

% Recommendation system is an important way to address the information overload problem, aiming to select the top-n items for users from a massive-scale item set. Modern recommendation system typically calls for the collaboration of representation
% learning and Approximate Nearest Neighbour search (ANNs). In the first place, users and items are
% represented by embeddings in the same latent space; in the second place, the item embeddings are
% organized with a specific ANNs index, like SCANN and HNSW, such that the top-n recommendation
% to the user can be efficiently accomplished. Although remarkable progress is achieved, the independence and incompatibility between representation models and ANN indexes impose limitations on end-to-end optimization~\cite{}, thereby constraining further advancements.

% Despite that the recommendation process is greatly
% accelerated from the above workflow, the recommendation quality will probably be restricted, given
% that the representation model is independently learned and can be incompatible with the ANNs index.

To address this limitation, many research efforts have been made. One prominent direction involves constructing tree-based matching indexes~\cite{zhu2018learning, zhu2019joint, feng2022recommender}, which
optimize both a matching model and a tree structure index for items. However, these methods often face challenges such as low inference efficiency due to the tree structure and limited utilization of item semantic information~\cite{feng2022recommender, si2023generative}. Recently, generative retrieval~\cite{GR_survey} has emerged as a promising new paradigm for information retrieval, which has been recently applied for generative recommendation~\cite{rajput2023recommender}. 
Unlike traditional representation-based user-item matching approaches, this paradigm employs an end-to-end generative model that predicts candidate item identifiers directly in an autoregressive manner. Specifically, Specifically, these methods begin by tokenizing each item $x$ into a set of discrete semantic codes\footnote{Note that we use "code" and "token" interchangeably.} $C=\{c_{1}, c_{2}, ...\}$, and then utilize an encoder-decoder model (e.g. Transformer~\cite{vaswani2017attention}) to serve as an end-to-end index for retrieval. In this setup, the encoder encodes the interaction history $\{x_{1}, x_{2}, ..., x_{t-1}\}$ between users and item, while the decoder predicts the code sequence $C_{t}$ of the next item $x_{t}$. The overall framework is illustrated in  ~\cref{fig:intro}.

However, existing generative recommendation approaches suffer from a significant drawback in how they utilize item information, often focusing narrowly on either behavioral or semantic aspects. Behavioral information is derived from user-item interaction histories, while semantic information encompasses textual or visual descriptions of items. For instance, RecForest~\cite{feng2022recommender} utilizes a pre-trained DIN model~\cite{zhou2018deep} to extract behavior-based item embeddings for constructing semantic codes, while TIGER utilizes the Sentence-T5 model~\cite{ni2022sentence} to derive semantic-based item embeddings from textual descriptions. However, these approaches often focus exclusively on one aspect, overlooking the complementary relationship between behavior and semantics. On one hand, advances in pre-trained modality encoders such as BERT~\cite{devlin2018bert} and ViT~\cite{dosovitskiy2020image} facilitate the integration of multimodal features, enhancing prior knowledge and finding wide applications in multimodal recommendation models~\cite{MMRec_Survey}. On the other hand, behavioral data captures user-specific preferences through interaction sequences, making it particularly effective in recommendation contexts. Conversely, semantic information offers broader, unbiased insights into item characteristics, fostering better generalization across different domains.

In this paper, we propose {\bf EAGER}, a novel two-str{\bf EA}m {\bf GE}nerati-ve {\bf R}ecommender with behavior-semantic collaboration. We analyze the challenges of modeling behavior and semantics within a unified generative framework and address them from the following three aspects:

Firstly, a {\bf unified generative architecture} for handling two distinct types of information is crucial. Given the inherent differences in feature spaces between behavior and semantics, directly integrating them through feature fusion at the encoder side poses challenges, as demonstrated in previous two-tower models~\cite{yuan2023go, shan2023beyond}. Therefore, our approach constructs separate codes for behavior and semantics, employing a two-stream generation architecture where each serves as a distinct supervision signal at the decoder side. This architecture includes a shared encoder for encoding user interaction history and two separate decoders for predicting behavior and semantic codes respectively, thereby avoiding premature feature interaction. During inference, we enhance the merging of results from both streams by utilizing the prediction entropy of each stream as a confidence measure for item ranking, ensuring effective predictions.

Secondly, ensuring {\bf sufficient and independent learning} is crucial to fully leverage the potential value of each type of information. Previous works~\cite{rajput2023recommender} have typically employed autoregressive approaches to learn each token one by one, focusing on discrete and local information rather than capturing global insights.
In EAGER, we introduce a global contrastive task with a summary token. This module draws inspiration from two main sources: (1) traditional dual-tower models use contrastive learning to acquire discriminative item features. Similarly, we aim for our decoder model to grasp global discriminatory capabilities alongside its autoregressive generation capability, thereby enhancing the extraction of item features within a contrastive paradigm; (2) Transformer models~\cite{devlin2018bert, dosovitskiy2020image} utilize special tokens to encapsulate global information, prompting us to append a summary token at the end of the token sequence. This token summarizes the accumulated knowledge in a unidirectional manner, serving as the focal point for distillation.

Thirdly, while separate decoding and prediction reranking have shown effectiveness, integrating {\bf subtle interaction} can enhance sharing of both knowledge flows. As mentioned earlier, direct feature-level interactions often yield sub-optimal outcomes~\cite{yuan2023go}. Therefore, we introduce a carefully crafted semantic-guided transfer task to promote implicit knowledge exchange. Specifically, we propose that semantic information can guide behavioral aspects, and we design an auxiliary transformer with dual objectives: reconstruction and recognition. The reconstruction objective involves predicting masked behavior tokens using global semantic features, while the recognition objective aims to differentiate whether a behavior token aligns with a specified global semantic feature. Through these objectives, this module indirectly optimizes interaction between behavioral and semantic features using the transformer model.

In summary, our main contributions are as follows:

\begin{itemize}[leftmargin=*]
    \item We introduce EAGER, a novel generative recommendation framework that integrates behavior and semantic information collaboratively.

    \item We propose a unified two-stream generative architecture, design a global contrastive module with a summary token to ensure sufficient and independent learning, and introduce a semantic-guided transfer module for subtle interaction.
    
    \item Extensive experiments on four public recommendation benchmarks demonstrate EAGER's superiority over existing methods, encompassing both generative and traditional paradigms.

\end{itemize}

\section{Related Work}

\vpara{Sequential Recommendation.} Using deep sequential models to capture user-item patterns in recommender systems has developed into a rich literature. GRU4REC~\cite{jannach2017recurrent} was the first to use GRU-based RNNs for sequential recommendations. SASRec~\cite{kang2018self} adopts self-attention which is similar to decoder-only transformer models. Inspired by the success of masked language modeling in language tasks, BERT4Rec~\cite{sun2019bert4rec} utilizes transformer models with masking strategies for sequential recommendation tasks. $S^3$-Rec~\cite{zhou2020s3} not only relies on masking but also pre-trains embeddings through four self-supervised tasks to enhance the quality of item and user embeddings. The above mentioned methods mainly depend on an approximate nearest neighbor (ANN) search index (e.g., Faiss~\cite{johnson2019billion}) to retrieve the next item.
Besides, tree-based methods~\cite{feng2022recommender, zhu2018learning, zhu2019joint} have shown
promising performance in recommender systems. For example, RecForest~\cite{feng2022recommender} constructs a forest by creating multiple trees and integrates a transformer-based structure for routing operations.
Recently, TIGER~\cite{rajput2023recommender} introduced the idea of semantic id, where each item is represented as a set of tokens derived from its side information, and then predicts the next item tokens in a seq2seq way.
In this work, we aim to further explore a two-stream generation architecture to act as an end-to-end index for top-k item retrieval.
\begin{figure*}[t]
  \centering 
  \includegraphics[width=0.9\linewidth]{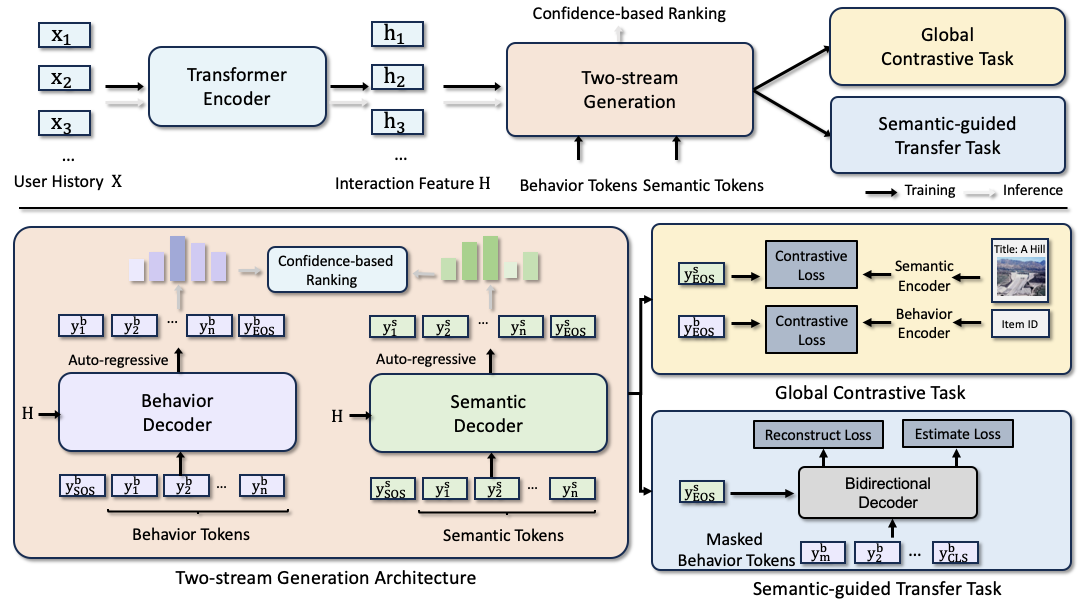}
  \caption{{\bf An overview of EAGER. EAGER consists of three major components: Two-Stream Generation Architecture (TSG), Global Contrastive Task (GCT), and Semantic-guided Transfer Task (STT).} }
  \label{fig:framework}
\end{figure*}

\vpara{Generative Retrieval.} Generative retrieval~\cite{GR_survey} has been recently proposed as a new retrieval paradigm, which consists of two main phases: discrete semantic tokenization~\cite{UIST,CoST,LMINDEXER} and autoregressive sequence generation~\cite{tay2022transformer,wang2022neural,rajput2023recommender}. In the domain of document retrieval, researchers have explored the use of pre-trained language models to generate diverse types of document identifiers. Notably, DSI~\cite{tay2022transformer} and NCI~\cite{wang2022neural} leverage the T5~\cite{raffel2020exploring} model to produce hierarchical document IDs. Conversely, SEAL~\cite{bevilacqua2022autoregressive} (with BART~\cite{lewis2020bart} backbone) and ULTRON~\cite{zhou2022ultron} (using T5) utilize titles or substrings as identifiers. Another approach, AutoTSG~\cite{zhang2023term}, adopts term-sets for identification purposes. Generative document retrieval has also extended to various domains. For instance, IRGen~\cite{zhang2023irgen} employs a ViT-based model for image search, while TIGER~\cite{rajput2023recommender} utilizes T5 for recommender systems. However, these studies often face challenges in large-scale item retrieval within recommender systems due to the resource-intensive nature of pre-trained language models. In contrast, our paper delves into integrating both behavior and semantics for generative retrieval in such systems.
%In document retrieval, researchers have investigated using pre-trained language models to generate 

\section{Method}

\subsection{Problem Formulation}

Given the entire set of items $\mathcal{X}$ and interacted items history ${\bf X} = \{{\bf x}_1, {\bf x}_2, \cdots, {\bf x}_{t-1} \} \in \mathcal{X}$ of a user, the sequence recommendation system returns a list of item candidates for the next item ${\bf x}_{t}$.

In generative recommendation, the identifier of each item ${\bf x}$ is represented as a serialized code ${\bf Y}=[{\bf y}_1, {\bf y}_2, \cdots, {\bf y}_l]\in \mathcal{Y}$, where $l$ is the length of the code. The goal of the generative model is learning a mapping $f: \mathcal{X}\rightarrow \mathcal{Y}$, which takes a user's interacted item sequence as input and generates item codes (candidate identifiers). For training, the model first feeds the user's behavior ${\bf X}$ into the encoder, then leverages the auto-regressive decoder to generate the item code ${\bf Y}$ step by step. The probability of interaction can be calculated by:
\begin{equation}
   p({\bf Y}|{\bf X}) = {\prod}_{i=1}^{l} p({\bf y}_i|\bf{X},{\bf y}_1,{\bf y}_2,\dots,{\bf y}_{i-1})
\end{equation}
During inference, the decoder performs beam search over the sequential codes when selecting top-n candidates.

\subsection{Overall Pipeline}

We present our overall EAGER framework in \cref{fig:framework}. EAGER consists of (1) a two-stream generation architecture to unify item recommendation for both behavior and semantic information, (2) a global contrastive task with a summary token to capture global knowledge for better auto-regressive generation quality, and (3) a semantic-guided transfer task to achieve the cross-information and cross-decoder interaction.  

First, in our two-stream generation architecture, we model user interactions history and obtain interaction features via the encoder. Then we extract both behavior and semantic features to construct two codes, and leverage two decoders to separately predict them in an auto-regressive way. Meanwhile, we optimize a summary token in our global contrastive task and leverage it to improve cross-decoder interaction in our semantic-guided transfer task. After training, we adopt a confidence-based ranking strategy to merge the results from two different predictions.

\subsection{Two-stream Generation Architecture}
To handle two different information, i.e. behavior and semantic, we leverage the powerful modeling capabilities of transformer models and design a two-stream generation architecture. This framework consists of a shared encoder for modeling user interaction, two separate codes and decoders for two-stream generation.

\noindent {\bf Shared Encoder.}  The sequence modeling of user interaction history ${\bf X} = \{{\bf x}_1, {\bf x}_2, \cdots\}$ is based on the stacked multi-head self-attention layers and feed-forward layers, as proposed in Transformer. It is worth nothing that we only adopt a shared encoder instead of two encoders, which is enough to generate rich representation for the subsequent separate decoding. We denote the encoded historical interaction features as ${\bf H}={\rm Encoder}({\bf X})$.

% The shared encoder is responsible for encoding the input item interaction sequence, extracting meaningful representations. By leveraging the multi-modal processing capabilities of the Transformer model, it can generate a rich representation for the subsequent separate decoders of different features.

\begin{figure}[t]
  \centering 
  \includegraphics[width=\linewidth]{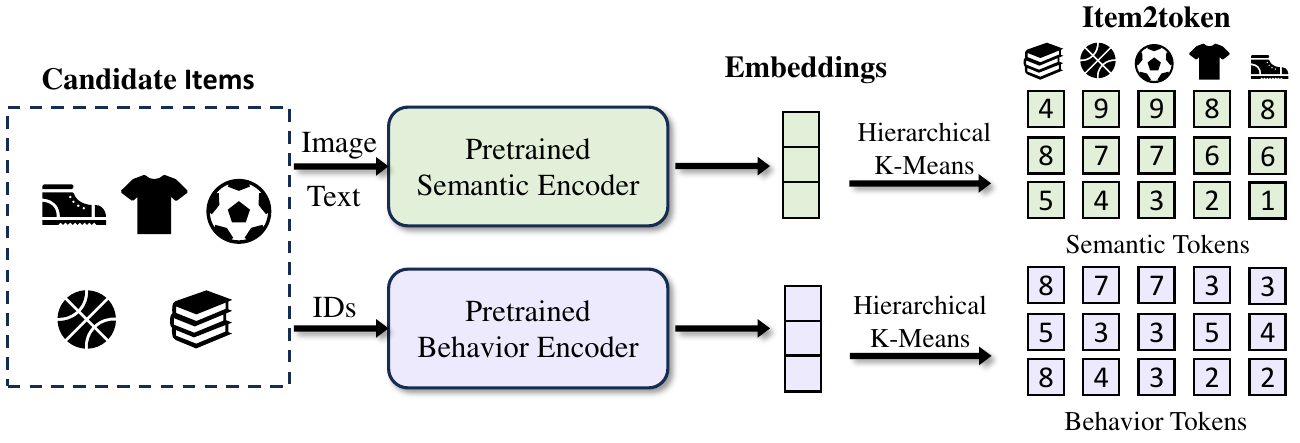}
  \caption{{\bf The illustration of dual codes.} }
  \label{fig:item2code}
  \vspace{-0.5cm}
\end{figure}

\noindent {\bf Dual Codes.} We first extract the behavior and semantic item embeddings ${\bf E}^{b}$ 
and ${\bf E}^{s}$ using two pre-trained models, where the behavior encoder is a two-tower model (e.g. DIN~\cite{zhou2018deep}) that only uses ID sequence for recommendation and the semantic encoder is a general modality representation model (e.g. Sentence-T5). With the two extracted embeddings, we separately apply the widely-used hierarchical k-means algorithm to each one, where each cluster is evenly divided into K child clusters until each child cluster merely contains one single item. As shown in \cref{fig:item2code}, we can obtain two codes ${\bf Y}^{b}$ and ${\bf Y}^{s}$, corresponding to behavior and semantic, respectively. 
% The detailed code construction process is listed in Appendix A.1.

\noindent {\bf Dual Decoders.} To accommodate two different codes, we employ two separate decoders to decode and generate the prediction for each of them, allowing each decoder to specialize in one single code. Compared to one shared decoder that generates two identifiers in an auto-regressive way, such design mitigates the supervision difference and offers higher efficiency with parallel generation. For training, we add a start token ${\bf y}_{\rm SOS}$ at the beginning of codes ${\bf Y}$ to construct the decoder inputs ${\bf \bar Y}=\{{\bf y}_{\rm SOS}, {\bf y}_1, {\bf y}_2, \cdots, {\bf y}_l\}$ 
%We obtain the output hidden states $\{{\bf z}_{\rm SOS}, {\bf z}_1, {\bf z}_2, \cdots, {\bf z}_l\}$  
%We obtain the output hidden states $\{{y}_{\rm SOS}, {y}_1, {y}_2, \cdots, {y}_l\}$ 
and utilize cross-entropy loss for prediction. The overall loss $\mathcal{L}_{\rm gen}$ is the sum of two generations losses $\mathcal{L}^{b}_{\rm gen}$ and $\mathcal{L}^{s}_{\rm gen}$, where each is given by:
\begin{equation}
   \mathcal{L}_{\rm gen}^{t} = {\sum}_{i=1}^{l} \log p({\rm y}^{t}_i|{\bf x}^{t},{\bf y}^{t}_{\rm SOS}, {\bf y}^{t}_1,\dots,{\bf y}^{t}_{i-1}), \quad t\in \{b, s\}
\end{equation}

% For inference, we generate two candidate identifiers parallelly. We adopt the ensemble strategy to first compute the logarithmic probability in each decoder and then sum logarithmic probabilities over all trees as the score of the item.  Afterwards, we can select the top-n items for recommendation.

\subsection{Global Contrastive Task}\label{sec:gct}

To endow each generative decoder with a sufficient discriminative capability, we design a global contrastive task with a summary token to distill the global knowledge.

\noindent {\bf Summary Token.} For the input ${\bf \bar Y}$ of each decoder, we consider the left-to-right order of auto-regressive generation and insert a learnable token ${\bf y}_{[{\rm EOS}]}$ at the end of the sequence to construct modified inputs ${\bf \tilde Y}=\{{\bf y}_{\rm SOS}, {\bf y}_1, {\bf y}_2, \cdots, {\bf y}_l, {\bf y}_{[{\rm EOS}]}\}$. This design encourages the preceding tokens in the codes to learn more comprehensive and discriminative knowledge, enabling the final token to make a summary. During updates, the gradient on the summary token can be backpropagated to the preceding tokens.

\noindent {\bf Contrastive Distillation.} To make the summary token capture global information, we adopt contrastive learning paradigm to distill the item embedding ${\bf E}^{b}$ 
and ${\bf E}^{s}$ from the pre-trained encoder. Here we adopt the positive-only contrastive metric~\cite{chen2021exploring} instead of commonly used Info-NCE~\cite{chen2020simple} to achieve this objective. The full loss $ \mathcal{L}_{\rm{con}}$ is given by summing two losses $ \mathcal{L}^{b}_{\rm{con}}$ and $ \mathcal{L}^{s}_{\rm{con}}$, where each is given by:
% \begin{equation}
%     \mathcal{L}^{t}_{\rm{con}} = {\mathcal{F}}(z^{t}_{[\rm EOS]}, {\bf E}^{t}), \quad t\in \{b, s\}
% \end{equation}
\begin{equation}
    \mathcal{L}^{t}_{\rm{con}} = {\mathcal{F}}(y^{t}_{[\rm EOS]}, {\bf E}^{t}), \quad t\in \{b, s\}
\end{equation}
where ${y}^{t}_{[\rm EOS]}$ corresponds to the embedding of the summary token ${\bf y}^{t}_{[\rm EOS]}$ and ${\mathcal{F}}(\cdot,\cdot)$ is the metric function, e.g. Smooth $\ell_1$.

\subsection{Semantic-guided Transfer Task}

Through the aforementioned components, our model can effectively utilize two types of information for prediction. However, we do not stop at this point. Instead of completely independent decoding, we further propose a semantic-guided transfer task to utilize the semantic knowledge to guide the behavior learning. 

To enable the knowledge flow between two sides while avoiding direct interaction, we build an independent bidirectional Transformer decoder as an auxiliary module. We first add a token ${\bf y}_{[\rm cls]}$ at the beginning of behavior codes to obtain the sequence $\{{\bf y}^{b}_{[\rm cls]}, {\bf y}^{b}_1, {\bf y}^{b}_2,\\ \cdots, {\bf y}^{b}_l\}$ as the input to the decoder. Then the embedding ${y}^{s}_{\rm EOS}$ of the semantic summary token ${\bf y}^{s}_{[\rm EOS]}$ is input to the cross attention, allowing each behavior token in the decoder to attend over global feature of the semantic. We denote the output features as $\{{\bf r}_{[\rm cls]}, {\bf r}_{1}, {\bf r}_{2}, \cdots, {\bf r}_{l}\}$. To conduct our transfer training, we design two  following objectives: reconstruction and recognition.

{\noindent \bf Reconstruction.} We reconstruct the masked behavior codes via the semantic global feature,  {{aiming to enable the each behavior token benefit from the semantic}}. For reconstruction training, we randomly mask m\% of tokens in the behavior code to obtain the masked code $\{{\bf y}^{b}_{[\rm cls]}, {\bf y}^{b}_{1}, {\bf y}^{b}_{[\rm  mask]}, \cdots, {\bf y}^{b}_{l}\}$, where ${\bf y}^{b}_{[\rm mask]}$ means the masked token. Then, we obtain the corresponding output features $\{{\bf r}_{[\rm cls]}, {\bf r}_{1}, {\bf r}_{[\rm mask]}, \cdots, {\bf r}_{l}\}$ and apply the contrastive loss to develop the reconstruction by:
\begin{equation}
\begin{aligned}
&{\mathcal L}_{i}  = \log \frac{ {\rm exp}({\bf r}^{b+}_{[{m}_i]} \cdot {\bf y}_i)  }{{\rm exp}({\bf r}^{b+}_{[{m}_i]} \cdot {\bf y}_i)  + \sum_{j=1}^{J} {\rm exp}({\bf r}^{b+}_{[{m}_i]} \cdot {\bf y}_j) }, \\
&{\mathcal L}_{\rm recon}  = -\frac{2}{N}\sum_{i=1}^{N/2}{\mathcal L}_{i}^b
\end{aligned}
\end{equation} %\in \mathbb{R}^{d_x}%
where ${\bf y}_i$ is the feature of ground truth of the $i$-th masked token,  ${\bf y}_j$ is the feature of sampled tokens.

{\noindent \bf Recognition.} Besides, we also build a binary classifier to judge whether the behavior codes is relevant or irrelevant to the semantic global feature. For recognition training, we construct negative samples by randomly replacing the m\% of tokens in the behavior code with the sampled irrelevant token, e.g. [23, 123, 32] $\rightarrow$ [23, 145, 32]. We add a linear layer on the corresponding output of the token [CLS] and utilize a linear layer with sigmoid activation to calculate the score $s^{+}/s^{-}$ for positive/negative samples. The binary cross-entropy loss is utilized for recognition, given by:
\begin{equation}
{\mathcal L}_{\rm recog}  = -{\rm log}(s^{+}) + {\rm log}(1-s^{-})
\end{equation}

\subsection{Training and Inference}

\noindent {\bf Training.} We combine the generation, contrastive, reconstruction and recognition losses to train our model, given by:
\begin{equation}
{\mathcal L}_{\rm EAGER}  = {\mathcal L}_{\rm gen} + \lambda_{1}{\mathcal L}_{\rm con} + \lambda_{2}({\mathcal L}_{\rm recon} +  {\mathcal L}_{\rm recog})
\end{equation}
where $\lambda_{1}$ and $\lambda_{2}$ are loss coefficients.

\noindent {\bf Inference with Confidence-based Ranking.} Since we have two results derived from the behavior and the semantic streams, we first obtain top-$k$ predictions via beam search from each stream. With the 2*$k$ prediction codes, we calculate the log probabilities over the codes as the confidence score of each prediction, which is similar to the perplexity used in the language model and the lower value indicates more confidence. Finally, we rank these predictions by their confidence scores and obtain the top-$k$ predictions, which corresponds to $k$ items. 

%% file: part/xjh.tex
\begin{table}[tp]
\centering
\caption{Statistics of the Datasets.}
\label{tab:dataset}
\resizebox{\linewidth}{!}{
\begin{tabular}{ccccc}
\toprule
Dataset & \#Users & \#Items & \#Interactions & \#Density \\ \midrule
Beauty &22,363  & 12,101  & 198,360 & 0.00073\\
Sports and Outdoors &35,598  & 18,357  &296,175 & 0.00045\\
Toys and Games &19,412  & 11,924 & 167,526 &0.00073 \\
Yelp & 30,431 & 20,033 & 316,354 & 0.00051 \\
%Gowalla & 29,858 & 40,988 & 1,027,464 &  0.00084\\
\bottomrule
\end{tabular}
}
\end{table}

\begin{table*}[htp]
\centering
\caption{Performance comparison of different methods. The best performance is highlighted in bold while the second best performance is underlined. The last column indicates the improvements over the best baseline models and all the results of Eager are statistically significant with p < 0.05 compared to the best baseline models.}
\begin{tabular}{clccccccccccc}
\toprule
\multirow{2}{*}{Dataset} & \multirow{2}{*}{Metric} & \multicolumn{3}{c}{\emph{Traditional}} & \multicolumn{3}{c}{\emph{Transformer-based}} & \multicolumn{2}{c}{\emph{Tree-based}} & \multicolumn{2}{c}{\emph{Generative}} & \multirow{2}{*}{Improv.} \\ \cmidrule(lr){3-5} \cmidrule(lr){6-8} \cmidrule(lr){9-10} \cmidrule(lr){11-12}
 &  & GRU4REC & Caser & HGN & SASRec & Bert4Rec & S\textasciicircum{}3-Rec & TDM & Recforest & TIGER & EAGER &  \\ \midrule
\multirow{6}{*}{Beauty} & Recall@5 & 0.0164 & 0.0205 & 0.0325 & 0.0387 & 0.0203 & 0.0387 & 0.0442 & \underline{0.0470}$^{*}$ & 0.0454 & \textbf{0.0618} & \cellcolor{linecolor}{31.49\%} \\
 & Recall@10 & 0.0283 & 0.0347 & 0.0512 & 0.0605 & 0.0347 & 0.0647 & 0.0638 & \underline{0.0664}$^{*}$ & 0.0648 & \textbf{0.0836} & \cellcolor{linecolor}{25.90\%} \\
 & Recall@20 & 0.0479 & 0.0556 & 0.0773 & 0.0902 & 0.0599 & \underline{0.0994} & 0.0876 & 0.0915$^{*}$ & - & \textbf{0.1124} & \cellcolor{linecolor2}{13.08\%} \\
 & NDCG@5 & 0.0099 & 0.0131 & 0.0206 & 0.0249 & 0.0124 & 0.0244 & 0.0323 & \underline{0.0341}$^{*}$ & 0.0321 & \textbf{0.0451} & \cellcolor{linecolor}{32.26\%} \\
 & NDCG@10 & 0.0137 & 0.0176 & 0.0266 & 0.0318 & 0.0170 & 0.0327 & 0.0376 & \underline{0.0400}$^{*}$ & 0.0384 & \textbf{0.0525} & \cellcolor{linecolor}{31.25\%} \\
 & NDCG@20 & 0.0187 & 0.0229 & 0.0332 & 0.0394 & 0.0233 & 0.0414 & 0.0438 & \underline{0.0464}$^{*}$ & - & \textbf{0.0599} & \cellcolor{linecolor}{29.09\%} \\ \midrule
\multirow{6}{*}{Sports} & Recall@5 & 0.0129 & 0.0116 & 0.0189 & 0.0233 & 0.0115 & 0.0251 & 0.0127 & 0.0149$^{*}$ & \underline{0.0264} & \textbf{0.0281} & \cellcolor{linecolor1}{6.44\%} \\
 & Recall@10 & 0.0204 & 0.0194 & 0.0313 & 0.0350 & 0.0191 & 0.0385 & 0.0221 & 0.0247$^{*}$ & \underline{0.0400} & \textbf{0.0441} & \cellcolor{linecolor2}{10.25\%} \\
 & Recall@20 & 0.0333 & 0.0314 & 0.0477 & 0.0507 & 0.0315 & \underline{0.0607} & 0.0349 & 0.0375$^{*}$ & - & \textbf{0.0659} & \cellcolor{linecolor1}{8.57\%} \\
 & NDCG@5 & 0.0086 & 0.0072 & 0.0120 & 0.0154 & 0.0075 & 0.0161 & 0.0096 & 0.0101$^{*}$ & \underline{0.0181} & \textbf{0.0184} & \cellcolor{linecolor1}{1.66\%} \\
 & NDCG@10 & 0.0110 & 0.0097 & 0.0159 & 0.0192 & 0.0099 & 0.0204 & 0.0110 & 0.0133$^{*}$ & \underline{0.0225} & \textbf{0.0236} & \cellcolor{linecolor1}{4.89\%} \\
 & NDCG@20 & 0.0142 & 0.0126 & 0.0201 & 0.0231 & 0.0130 & \underline{0.0260} & 0.0141 & 0.0164$^{*}$ & - & \textbf{0.0291} & \cellcolor{linecolor2}{11.92\%} \\ \midrule
\multirow{6}{*}{Toys} & Recall@5 & 0.0097 & 0.0166 & 0.0321 & 0.0463 & 0.0116 & 0.0443 & 0.0305 & 0.0313$^{*}$ & \underline{0.0521} & \textbf{0.0584} & \cellcolor{linecolor2}{12.09\%} \\
 & Recall@10 & 0.0176 & 0.0270 & 0.0497 & 0.0675 & 0.0203 & 0.0700 & 0.0359 & 0.0383$^{*}$ & \underline{0.0712} & \textbf{0.0714} & \cellcolor{linecolor1}{0.28\%} \\
 & Recall@20 & 0.0301 & 0.0420 & 0.0716 & 0.0941 & 0.0358 & \textbf{0.1065} & 0.0442 & 0.0483$^{*}$ & - & \underline{0.1024} & -3.85\% \\
 & NDCG@5 & 0.0059 & 0.0107 & 0.0221 & 0.0306 & 0.0071 & 0.0294 & 0.0214 & 0.0260$^{*}$ & \underline{0.0371} & \textbf{0.0464} & \cellcolor{linecolor}{25.07\%} \\
 & NDCG@10 & 0.0084 & 0.0141 & 0.0277 & 0.0374 & 0.0099 & 0.0376 & 0.0230 & 0.0285$^{*}$ & \underline{0.0432} & \textbf{0.0505} & \cellcolor{linecolor2}{16.90\%} \\
 & NDCG@20 & 0.0116 & 0.0179 & 0.0332 & 0.0441 & 0.0138 & \underline{0.0468} & 0.0284 & 0.0310$^{*}$ & - & \textbf{0.0538} & \cellcolor{linecolor2}{14.96\%} \\ \midrule
\multirow{6}{*}{Yelp} & Recall@5 & 0.0152 & 0.0151 & 0.0186 & 0.0162 & 0.0051 & 0.0201 & 0.0181 & \underline{0.0220}$^{*}$ & 0.0212$^{*}$ & \textbf{0.0265} & \cellcolor{linecolor}{20.45\%} \\
 & Recall@10 & 0.0263 & 0.0253 & 0.0326 & 0.0274 & 0.0090 & 0.0341 & 0.0287 & 0.0302$^{*}$ & \underline{0.0367}$^{*}$ & \textbf{0.0453} & \cellcolor{linecolor2}{12.69\%} \\
 & Recall@20 & 0.0439 & 0.0422 & 0.0535 & 0.0457 & 0.0161 & \underline{0.0573} & 0.0422 & 0.0449$^{*}$ & {0.0552}$^{*}$ & \textbf{0.0724} & \cellcolor{linecolor2}{11.56\%} \\
 & NDCG@5 & 0.0099 & 0.0096 & 0.0115 & 0.0100 & 0.0033 & 0.0123 & 0.0121 & 0.0119$^{*}$ & \underline{0.0146}$^{*}$ & \textbf{0.0177} & \cellcolor{linecolor1}{3.51\%} \\
 & NDCG@10 & 0.0134 & 0.0129 & 0.0159 & 0.0136 & 0.0045 & 0.0168 & 0.0154 & 0.0163$^{*}$ & \underline{0.0194}$^{*}$ & \textbf{0.0242} & \cellcolor{linecolor2}{18.63\%} \\
 & NDCG@20 & 0.0178 & 0.0171 & 0.0212 & 0.0182 & 0.0063 & 0.0226 & 0.0208 & 0.0210$^{*}$ & \underline{0.0230}$^{*}$ & \textbf{0.0311} & \cellcolor{linecolor2}{19.62\%} \\ \bottomrule
\end{tabular}\label{tab:results}
\end{table*}

\section{Experiments}
We analyze the proposed EAGER method and demonstrate its effectiveness by answering the following research questions:

\begin{itemize}[leftmargin=*]
	\item {RQ1}: How does EAGER perform compared with existing best-performing sequential recommendation methods among different datasets?
	\item {RQ2}: Do two-stream generation architecture, global contrastive task module, and semantic-guided transfer task module all contribute to the effectiveness of EAGER?
	\item {RQ3}: How do different ablation variants and hyper-parameter settings affect the performance of EAGER?
%	\item {RQ4}: Does DisCover disentangle the version-variant factors?
\end{itemize}

\subsection{Experimental Setting}
% \subsubsection{Dataset}
\vpara{ Dataset. } We conduct experiments on four open-source datasets commonly used in the sequential recommendation task. For all datasets, we group the interaction records by users and sort them by the interaction timestamps ascendingly. Following~\cite{rendle2010factorizing, zhang2019feature}, we only keep the 5-core dataset, which filters unpopular items and inactive users with fewer than five interaction records. Statistics of these datasets are shown in Table \ref{tab:dataset}.
\begin{itemize}[leftmargin=*]
	% \item {Beauty}: .
 %    \item {Sports and Outdoors}: 
 %    \item {Toys and Games}:
    \item \textbf{Amazon}: Amazon Product Reviews dataset~\cite{mcauley2015image}, containing user reviews and item metadata from May 1996 to July 2018. Here we use three categories ({Beauty}, {Sports and Outdoors}, and {Toys and Games}) for evaluations.
	% \item {Gowalla}: A widely used check-in dataset \cite{liang2016modeling} from the Gowalla platform. Similarly, we use the 10-core setting \cite{he2016vbpr}.
	\item \textbf{Yelp 2019\footnote{https://www.yelp.com/dataset}}: Yelp Challenge releases the review data for small businesses (e.g., restaurants). Following the previous setting~\cite{zhou2020s3}, we only use the transaction records from \textit{January 1st, 2019} to \textit{December 31st, 2019}. We view these businesses as items. 

\end{itemize}

\begin{table*}[t]
\centering
\caption{Ablation studies by selectively discarding the Two-stream Generation Architecture (TSG), Global Contrastive Task (GCT), and semantic-guided Transfer Task (STT). We study  EAGER on different datasets to reveal the model-agnostic capability of the proposed modules.}
\setlength{\tabcolsep}{0.3em}
\begin{tabular}{ccccccccccccccc}
\toprule
\multicolumn{3}{c}{Variants} & \multicolumn{6}{c}{Beauty} & \multicolumn{6}{c}{Toys and Games} \\ \cmidrule(lr){1-3} \cmidrule(lr){4-9} \cmidrule(lr){10-15}
TSG & GCT & STT & R@5 & NDCG@5 & R@10 & NDCG@10 & R@20 & NDCG@20 & R@5 & NDCG@5 & R@10 & NDCG@10 & R@20 & NDCG@20 \\ \midrule
 &  &  & 0.0512 & 0.0370 & 0.0699 & 0.0430 & 0.0943 & 0.0491 & 0.0436 & 0.0344 & 0.0545 & 0.0379 & 0.0657 & 0.0407 \\
\rowcolor{linecolor1} \checkmark &  &  & 0.0582 & 0.0425 & 0.0795 & 0.0496 & 0.1034 & 0.0567 & 0.0526 & 0.0428 & 0.0646 & 0.0469 & 0.0879 & 0.0510 \\
\rowcolor{linecolor2} \checkmark & \checkmark &  & 0.0604 & 0.0439 & 0.0815 & 0.0514 & 0.1091 & 0.0587 & 0.0563 & 0.0454 & 0.0699 & 0.0497 & 0.0974 & 0.0525 \\
\rowcolor{linecolor} \checkmark & \checkmark & \checkmark & 0.0618 & 0.0451 & 0.0836 & 0.0525 & 0.1124 & 0.0599 & 0.0584 & 0.0464 & 0.0714 & 0.0505 & 0.1024 & 0.0538 \\ \bottomrule
\end{tabular}  \label{tab:ablation}
\end{table*}

% \subsubsection{Evaluation Metrics}
\noindent {\bf Evaluation Metrics. }
We employ two broadly used criteria for the matching phase, \ie, Recall and Normalized Discounted Cumulative Gain (NDCG). We report metrics computed on the top 5/10/20 recommended candidates. Following the standard evaluation protocol~\cite{kang2018self}, we use the leave-one-out strategy for evaluation. For each
item sequence, the last item is used for testing, the item before the last is used for validation, and the
rest is used for training. During training, we limit the number of items in a user’s history to 20.

% \subsubsection{Implementation Details}
\vpara{Implementation Details. }
For two-stream generation architecture, we set the number of encoder layers to 1, and the number of decoder layers to 4. Following previous works~\cite{feng2022recommender, rajput2023recommender}, we adopt pre-trained DIN as our behavior encoder and Sentence-T5 as our semantic encoder, and set the hidden size to 128 as reported in ~\cite{rajput2023recommender}. The cluster number $k$ in hierarchical k-means is set to 256. For global contrastive task, we adopt Smooth $\ell_1$ distance to serve as the distilling loss. For semantic-guided transfer task, we randomly mask 50\% behavior codes for reconstruction, and randomly replace 50\% behavior codes with the sampled code to construct negative pairs for recognition. To train our model, we adopt Adam optimizer with the learning rate 0.001, and employ the warmup strategy for stable training. EAGER is not sensitive to the hyper-parameters for the GCT and STT tasks because these tasks converge quickly. So the loss coefficients $\lambda_{1}, \lambda_{2}$ are both set to 1.

\subsection{Performance Comparison (RQ1)}

% \subsubsection{Comparison Baselines}
\vpara{Baselines. } The baseline methods chosen for comparison can be split into the following four categories: 

\noindent (1) For \emph{traditional sequential} methods, we have:
\begin{itemize}[leftmargin=*]
    % \item {Y-DNN} \cite{covington2016deep}: A successful industrial recommender that generates the user’s representation by pooling the embeddings of historically interacted items.
    \item \textbf{GRU4REC} \cite{hidasi2015session}:  An early attempt to introduce recurrent neural networks into recommendation.
    \item \textbf{Caser} \cite{tang2018personalized}: a CNN-based method capturing high-order Markov Chains by applying horizontal and vertical convolutional operations for sequential recommendation.
    \item \textbf{HGN} \cite{ma2019hierarchical}: it adopts hierarchical gating networks to capture long-term and short-term user interests. 
    % \item \textbf{FDSA} \cite{zhang2019feature}: Feature-level Deeper Self-Attention Network incorporates item features in addition to the item embeddings as part of the input sequence in the Transformers.
\end{itemize}
(2) For \emph{transformer-based} methods, we have:
\begin{itemize}[leftmargin=*]
    \item \textbf{SASRec} \cite{kang2018self}:  SASRec models user’s behavior with Transformer encoder, where multi-head attention mechanism is attached to great importance.
    \item \textbf{BERT4Rec} \cite{sun2019bert4rec}: it uses a cloze objective loss for sequential recommendation by the bidirectional self-attention mechanism.
    \item \textbf{S\textasciicircum{}3-Rec} \cite{zhou2020s3}: S\textasciicircum{}3-Rec proposes pre-training a bi-directional Transformer on self-supervision tasks to improve the sequential recommendation.
\end{itemize}
(3) For \emph{tree-based} methods, we have:
\begin{itemize}[leftmargin=*]
    \item \textbf{TDM} \cite{zhu2018learning}: TDM uses a tree index to organize items (each leaf node in the
tree corresponds to an item) and designs a maximum heap-based tree model for retrieval.
    \item \textbf{RecForest} \cite{feng2022recommender}: RecForest constructs a forest with multiple k-branch trees and integrates a transformer-based structure for routing operations.
\end{itemize}
(4) For \emph{generative} methods, we have:
\begin{itemize}[leftmargin=*]
    \item \textbf{TIGER} \cite{rajput2023recommender}: TIGER uses pretrained T5 to learn semantic ID for each item and autoregressively decodes the identifiers of the target candidates with semantic ID.
\end{itemize}
% Detailed baseline implementation can be found in  Appendix\wy{[TODO]}.

%Overall, the results across multiple evaluation metrics consistently indicate that
\vpara{Results. } Tab.~\ref{tab:results} reports the overall performance of four datasets. The results for all baselines without the superscript $^*$ are taken from the publicly accessible results~\cite{rajput2023recommender, zhou2020s3}. For missing statistics, we reimplement the baseline and report our experimental results. From the results, we have the following observations:
\begin{itemize}[leftmargin=*]
    \item \textbf{EAGER almost achieves better results than base models among different datasets.} Especially, EAGER performs considerably better on the Beauty benchmark compared to the second-best baseline with up to 31.49\% improvement in Recall@5 and 32.26\% improvement in NDCG@5 compared to TIGER. Similarly, on the larger Yelp dataset, EAGER is 20.45\% and 3.51\% better in Recall@5 and NDCG@5, respectively. We attribute the improvements to the fact that EAGER succeeds in integrating behavior and semantics information under a two-stream unified generative architecture with dual identifiers.
    \item \textbf{EAGER beats the previous generative models on most datasets.} EAGER differs from existing models through its two-stream decoder architecture and multi-task training, which facilitate a deeper understanding of behavior-semantic relationships and capture crucial global information of inter-codes. These superior improvements validate the effectiveness of our designs and the necessity of incorporating both behaviors and semantic information.

    \item \textbf{Generative models outperform other traditional baselines in most cases across four datasets.} The limitation could potentially arise from the utilization of a simplistic inner product matching approach, which may restrict their ability to effectively model intricate user-item interactions. Furthermore, in practical scenarios, the construction of ANN indexes primarily focuses on achieving rapid matching, leading to additional performance degradation due to misaligned optimization objectives. However, the challenge can be overcome by generative methods that leverage beam search strategies to directly predict item codes, thereby boosting the model's resilience and robustness. 

\end{itemize}

% \end{table*}

\subsection{Ablation Study (RQ2)}
%\subsubsection{Analysis of key building modules.}
We evaluated the performance impact of EAGER’s components via an ablation study. Specifically, we gradually discard the semantic-guided transfer task (STT), global constrastive task (GCT) and two-stream generation architecture (TSG) from EAGER to obtain ablation architectures. The results are reported in Table \ref{tab:ablation}, we can observe that:
\begin{itemize}[leftmargin=*]
	\item Removing any TSG, GCT or STT leads to performance degradation while removing all modules (\ie, the base model) leads to the worst performance among different datasets. These results demonstrate the effectiveness and robustness of the proposed three modules as well as the benefits of the two-stream generative paradigm. 
	\item Removing GCT leads to more performance drops than removing STT, suggesting that global information distillation of the inter-code is slightly more important than the knowledge flow between the intra-code. The observation also highlights the crucial role of both tasks in enabling the model to acquire more powerful dual item identifiers.
 %These results again verify the effectiveness of the.
    \item Removing TSG leads to the most performance decline, which indicates that the base model can significantly enhance its performance by integrating behavior-semantic information of items. The results again verify the superiority of the dual decoder architecture with confidence-based
    ranking.
\end{itemize}

\subsection{Model Analysis (RQ3)}

\begin{table}[tp]
\centering
\caption{Analysis of two-stream generation architecture.} 
\begin{tabular}{lcccc}
\toprule
Dataset   & \multicolumn{4}{c}{Yelp}         \\ \cmidrule{2-5} 
Information    & R@5    & NDCG@5 & R@10   & NDCG@10 \\ \midrule
Behav+Text    & 0.0265 & 0.0177 & 0.0453 & 0.0242  \\
Behav+Vis  & 0.0259 & 0.0171	& 0.0440 & 0.0236   \\
\rowcolor{linecolor} Behav+Text+Vis    & {\bf 0.0283} & {\bf 0.0187}	& {\bf 0.0484} & {\bf 0.0252}  \\ 
\bottomrule
\end{tabular}
\end{table}

\begin{table}[tp]
\centering
\caption{Analysis of globel token position in global contrastive task on different datasets.} 
\begin{tabular}{lcccc}
\toprule
Dataset   & \multicolumn{4}{c}{Beauty}         \\ \cmidrule{2-5} 
Type    & R@5    & NDCG@5 & R@10   & NDCG@10 \\ \midrule
Head    & 0.0473 &0.0337 &0.0612 &0.0401  \\
Mean  & 0.0559 & 0.0431	& 0.0760 & 0.0502   \\
\rowcolor{linecolor} Tail    & {\bf 0.0604} & {\bf 0.0439}	& {\bf 0.0815} & {\bf 0.0514}  \\ \midrule
Dataset   & \multicolumn{4}{c}{Toys and Games} \\ \cmidrule{2-5} 
Type    & R@5    & NDCG@5 & R@10   & NDCG@10 \\ \midrule
Head    & 0.0441 &0.0303 &0.0513 &0.0343  \\
Mean    & 0.0522 & 0.0406 & 0.0617 & 0.0446 \\
\rowcolor{linecolor} Tail    & {\bf 0.0563} & {\bf 0.0454} & {\bf 0.0699} & {\bf 0.0497} \\
\bottomrule
\end{tabular}
\end{table}

\begin{table}[tp]
\centering
\caption{Analysis of the inference speed (second per sample, topk=5, beam size=100) on Beauty dataset.} 
\begin{tabular}{lcccc}
\toprule
        & DIN    & RecForest & TIGER   & EAGER \\ \midrule
Speed    & 0.2349    & 0.0499 & 0.0281   & 0.0325 \\
Parameters    & 93 & 51 & 14 & 87  \\
\bottomrule
\end{tabular}
\end{table}

\vpara{Analysis of Two-stream Generation Architecture. } To further explore the role of our dual-stream structure in integrating behavior and semantics, we conduct experiments with visual features introduced on Yelp. We use ViT-B to extract image features of item covers, and then perform the same operation to get the visual-based semantic code. As shown in Table, we can see that: (1) The combination of textual semantic features and behavioral information is more effective than that of visual semantic features. This may be because text contains more intuitive and richer information than visual semantics, which can be too abstract and noisy. (2) Moreover, we construct a three-stream architecture to integrate textual semantics, visual semantics, and behavioral information at once, which results in improved performance. This phenomenon further reflects the effectiveness and generalizability of our architecture for seamless integration of different types of information.

\vpara{Analysis of Global Contrastive Task.} As analyzed in Section \ref{sec:gct}, we consider the auto-regressive decoder lacks discriminative capability. To tackle the issue, we design a additional global token to distill the global knowledge from pretrained behavior/semantic encoder. In this study, we focus on investigating the impact of different token types and contrastive metrics on the model's performance.

\begin{itemize}[leftmargin=*]

\item{{\bf Token Type.}} The design of the summary token plays a crucial role, as it represents global information while having an impact on the autoregressive generation of the code. We investigate three token types: 'Head' places the token at the beginning of the code, 'Tail' puts the token at the end of the code, and 'Mean' directly uses the mean of the code features. The results in Table indicate that 'Tail' achieves the best performance, which is consistent with our previous discussion. Placing the token at the beginning results in unstable training of autoregressive generation due to frequent updates, while averaging directly impacts the code features, causing conflict with generative training. Instead, placing the token at the end neither directly affects the generation of the preceding code, nor indirectly optimizes the code representation through gradient reversal.

\item{{\bf Contrastive metric.}} 
In the experiment, we evaluate the performance of three widely adopted distance metrics, e.g. Cosine, InfoNCE, and Smooth $\ell_1$, as loss functions. The results presented in Tab.~\ref{tab:metric} demonstrate that the positive-only contrastive metrics, specifically Cosine and Smooth $\ell_1$, outperform the commonly used InfoNCE. This superiority can be attributed to two main reasons: (1) the positive-only contrastive learning approach, combined with frozen pre-trained features, mitigates the risk of representation collapsing, as supported by prior works~\cite{chen2021exploring}. (2) the dataset often comprises items belonging to specific categories with limited semantic diversity, such as Books or Sports, which introduces noise and confusion for negative samples. These challenging negatives pose difficulty in effective mining, leading to suboptimal optimization results.

\end{itemize}

% \begin{itemize}[leftmargin=*]
%     \item  
%     \item  Metric function to serve as. Therefore a reliable metric function is vital for capturing global information. In this experiment, we select three commonly used distances (\eg\ cosine, infoNCE, and smooth l1) to serve as the . 
% \end{itemize}

\begin{table}[tp]
\centering
\caption{Analysis of metric functions in global contrastive task on different datasets.} 
\begin{tabular}{lcccc}
\toprule
Dataset   & \multicolumn{4}{c}{Beauty}         \\ \cmidrule{2-5} 
Metric    & R@5    & NDCG@5 & R@10   & NDCG@10 \\ \midrule
\rowcolor{linecolor} Cosine    & {\bf 0.0620} & {\bf 0.0458} & {\bf 0.0842} & {\bf 0.0535}  \\
InfoNCE   & 0.0611 & 0.0446 & 0.0820 & 0.0525  \\
Smooth $\ell_1$ & 0.0618 & 0.0451 & 0.0836 & 0.0525  \\ \midrule
Dataset   & \multicolumn{4}{c}{Toys and Games} \\ \cmidrule{2-5} 
Metric    & R@5    & NDCG@5 & R@10   & NDCG@10 \\ \midrule
Cosine    & 0.0578 & 0.0448 & 0.0686 & 0.0488  \\
InfoNCE   & 0.0558 & 0.0440 & 0.0663 & 0.0478  \\
\rowcolor{linecolor} Smooth $\ell_1$ & {\bf 0.0584} & {\bf 0.0464} & {\bf 0.0714} & {\bf 0.0505}  \\
\bottomrule
\end{tabular}\label{tab:metric}
\vspace{-0.5cm}
\end{table}

\begin{figure}[htp]
    \centering
    \begin{center}
    \begin{subfigure}{0.49\linewidth}
        \centering
        \includegraphics[width=\linewidth]{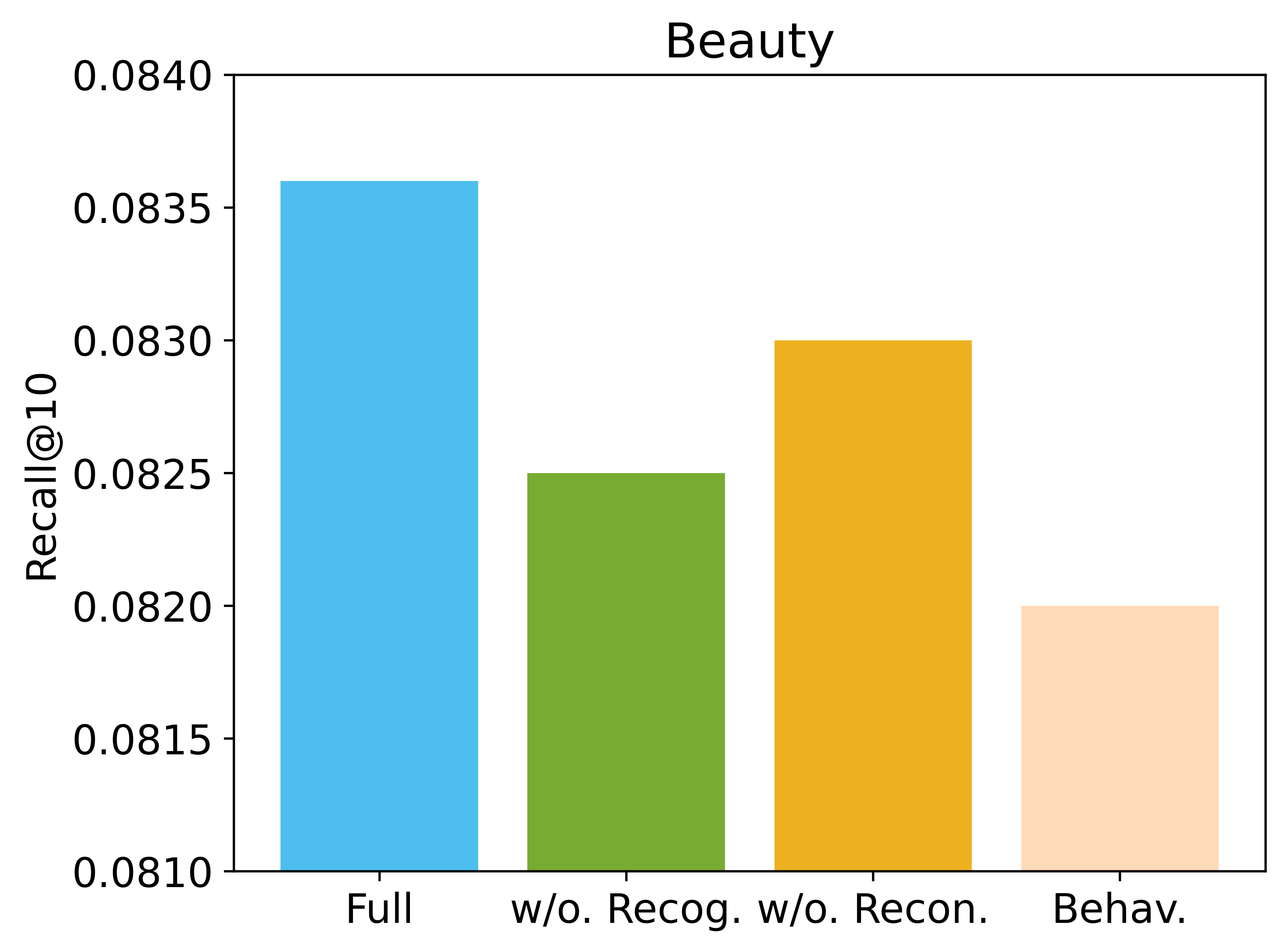}
        \caption{Beauty}
        \label{fig:ctt_base}
	\end{subfigure}
    \begin{subfigure}{0.49\linewidth}
        \centering
        \includegraphics[width=\linewidth]{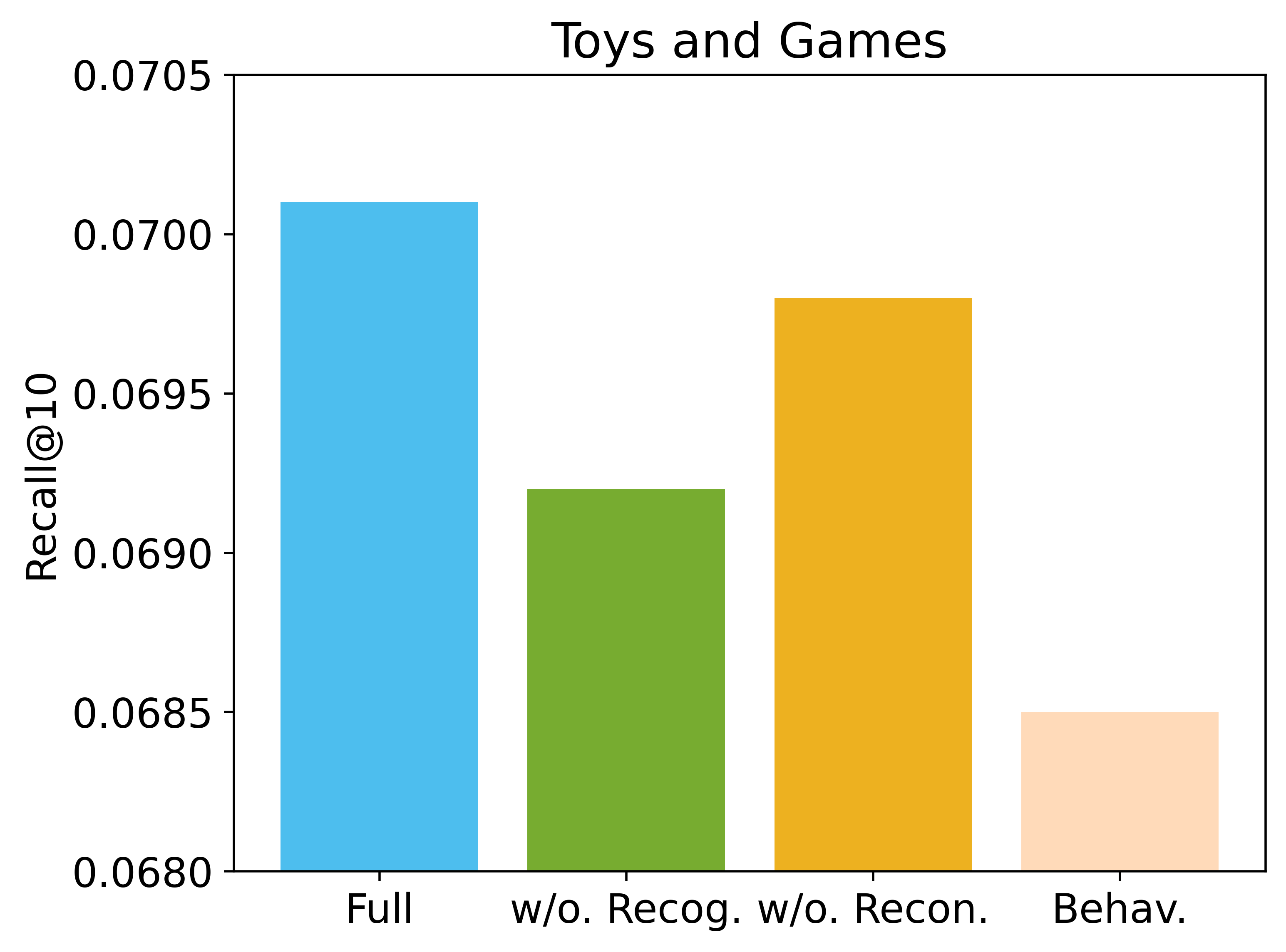}
        \caption{Toys and Games}
        \label{fig:ctt_ours}
	\end{subfigure}

    \caption{Analysis of semantic-guided transfer task module, where 'Behav.' means we swap the guidance direction.} %, where 'Behav.' means we swap the guidance direction.
    \label{fig:ctt}
    \end{center}
    \vspace{-0.5cm}
\end{figure}

\noindent {\bf Analysis of Semantic-guided Transfer Task. } We analyze the objectives and direction in our semantic-guided transfer task. 
\begin{itemize}[leftmargin=*]

    \item{{\bf Transfer Objective.}} We design two transfer objectives, i.e., reconstruction and recognition. As illustrated in Fig.~\ref{fig:ctt}, removing either objective leads to a certain degree of performance degradation, with recognition playing a more vital role. We hypothesize this is because the recognition task carries out more high-level knowledge transfer than the local-level reconstruction. 

    \item{{\bf Transfer Direction.}} Besides, we also investigate the behavior-guided transfer by swapping the guidance direction. It can be observed that the results is inferior to the semantic guidance, which is reasonable since the semantic information can provide more prior knowledge related to the item itself.

\end{itemize}

\subsection{Hyper-Parameter Analysis (RQ3)}

\begin{figure}[tp]
    \centering
    \begin{center}
    \begin{subfigure}{0.49\linewidth}
        \centering
        \includegraphics[width=\linewidth]{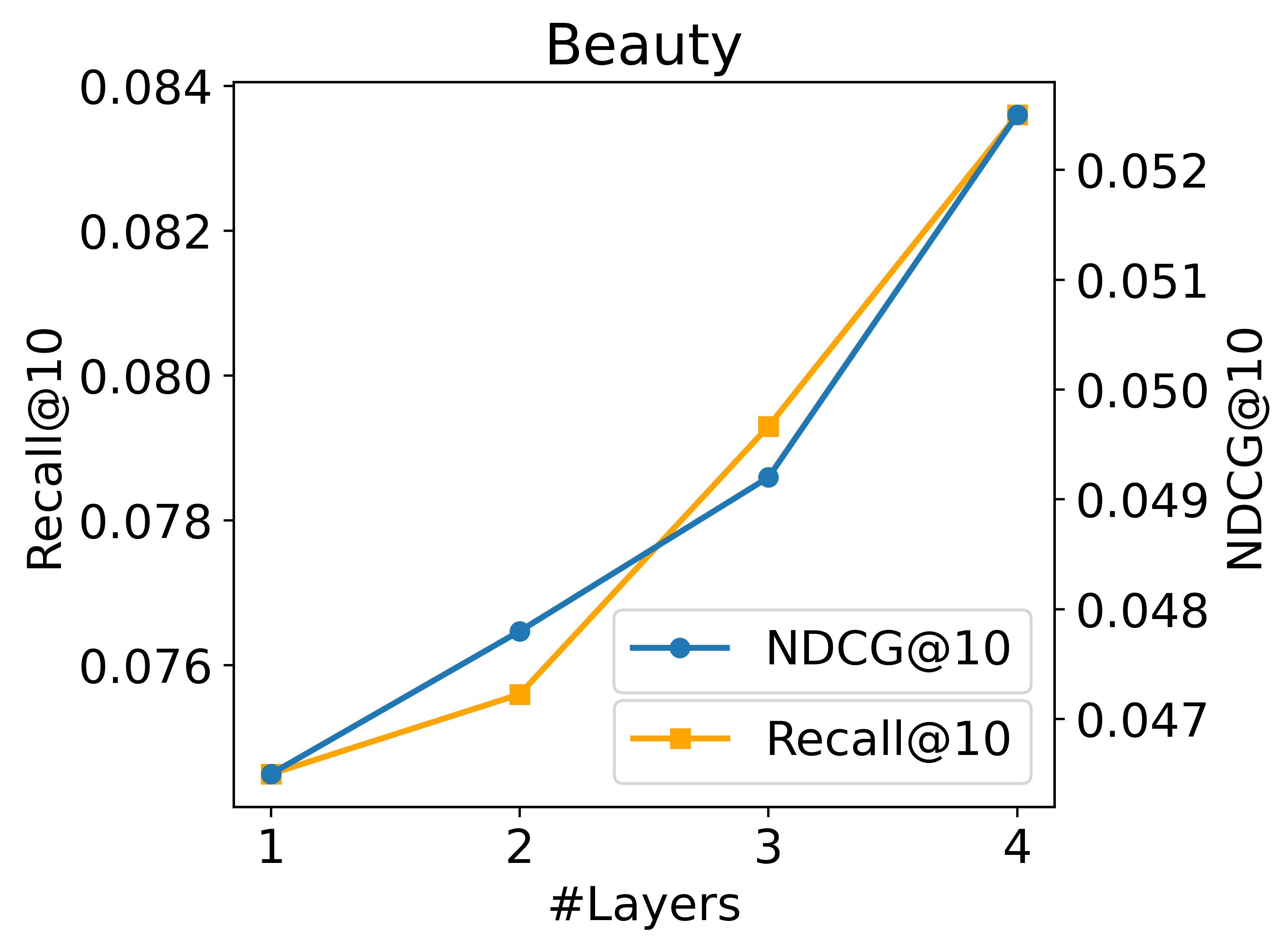}
        \caption{Beauty}
        \label{fig:layer_base}
	\end{subfigure}
    \begin{subfigure}{0.49\linewidth}
        \centering
        \includegraphics[width=\linewidth]{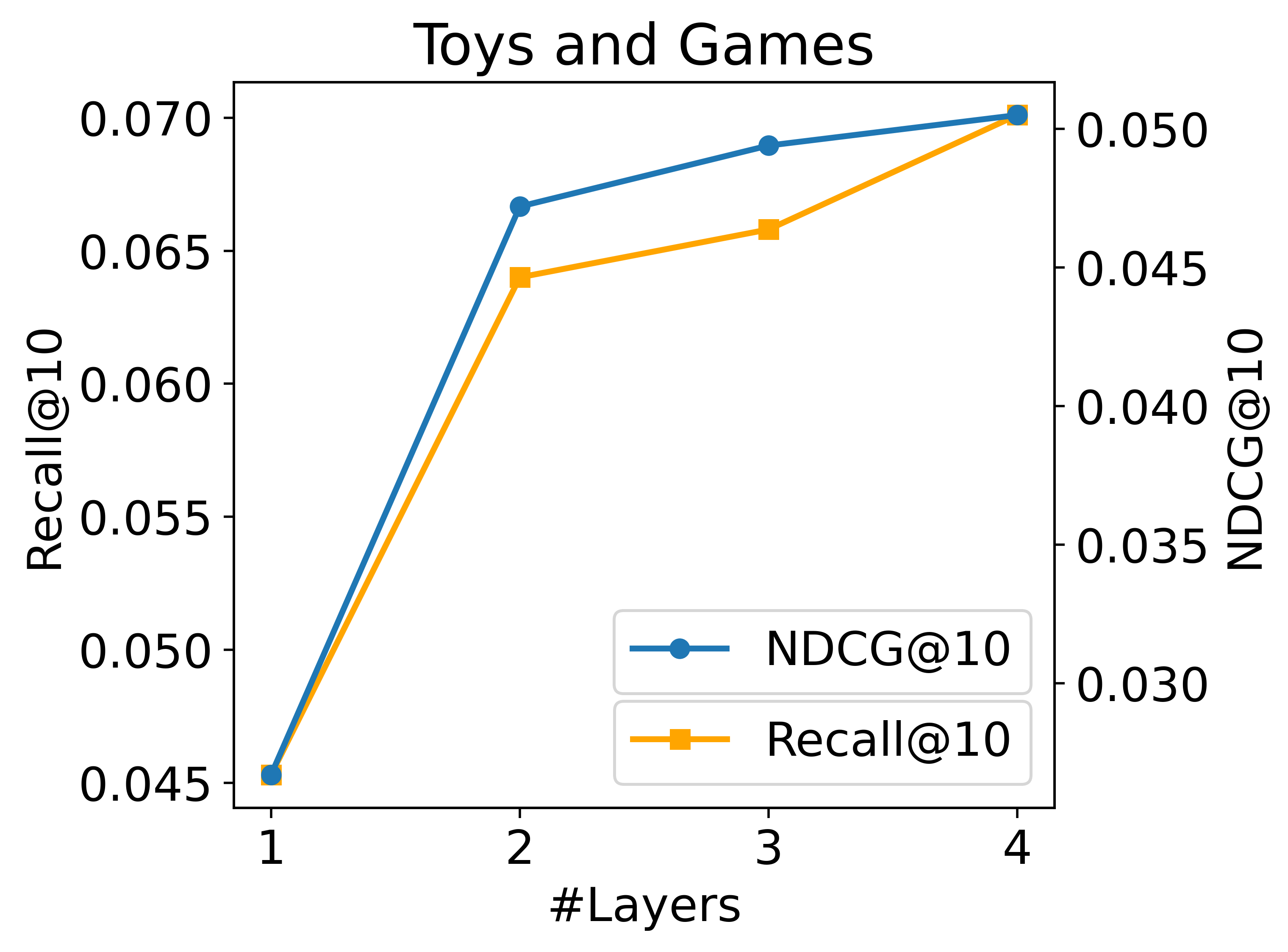}
        \caption{Toys and Games}
        \label{fig:layer_ours}
	\end{subfigure}

    \caption{Analysis of the number of transformer layers in decoders.}
    \label{fig:layer}
    \end{center}
    %\vspace{-0.5cm}
\end{figure}

\vpara{Layer Number.} In our practice, we found that the number of encoder layers has a negligible effect on performance, whereas the number of decoder layers has a more significant influence. It suggests that the decoder plays a more important role in our EAGER. Therefore, we focus on the decoder layer here. To study the impact of the number of decoder layers on model performance, we analyze the changes in Recall@10 and NDCG@10 across two datasets by varying layer scales. As shown in Fig. \ref{fig:layer}, there is a continuous improvement in the model’s performance in both datasets by increasing the number of layers. This can be attributed to the fact that larger parameters can enhance the model’s expressive capability. However, the deeper the model, the slower the inference speed.

\begin{figure}[tp]
    \centering
    \begin{center}
    \begin{subfigure}{0.49\linewidth}
        \centering
        \includegraphics[width=\linewidth]{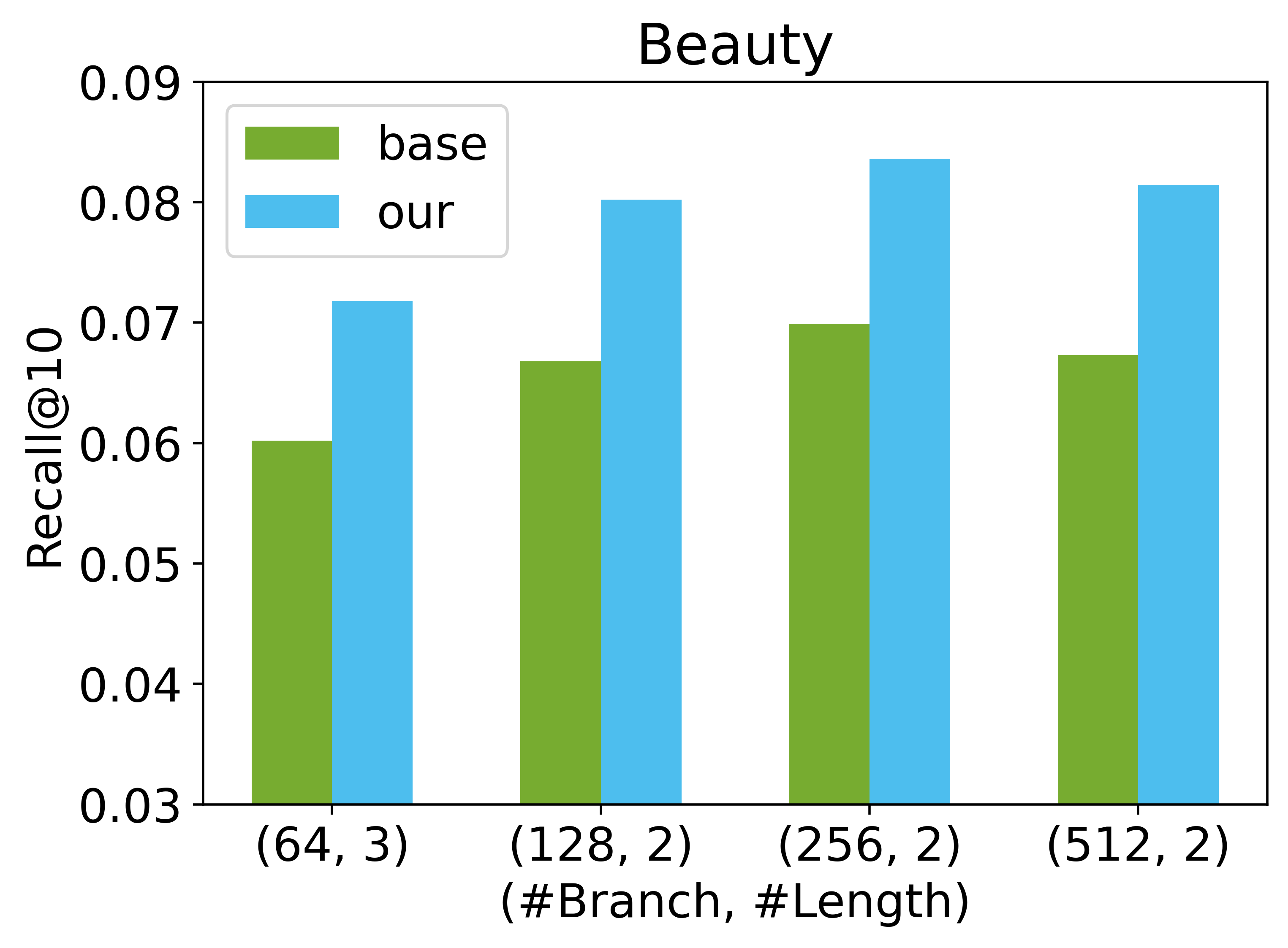}
        \caption{Beauty}
        \label{fig:cluster_beauty}
	\end{subfigure}
    \begin{subfigure}{0.49\linewidth}
        \centering
        \includegraphics[width=\linewidth]{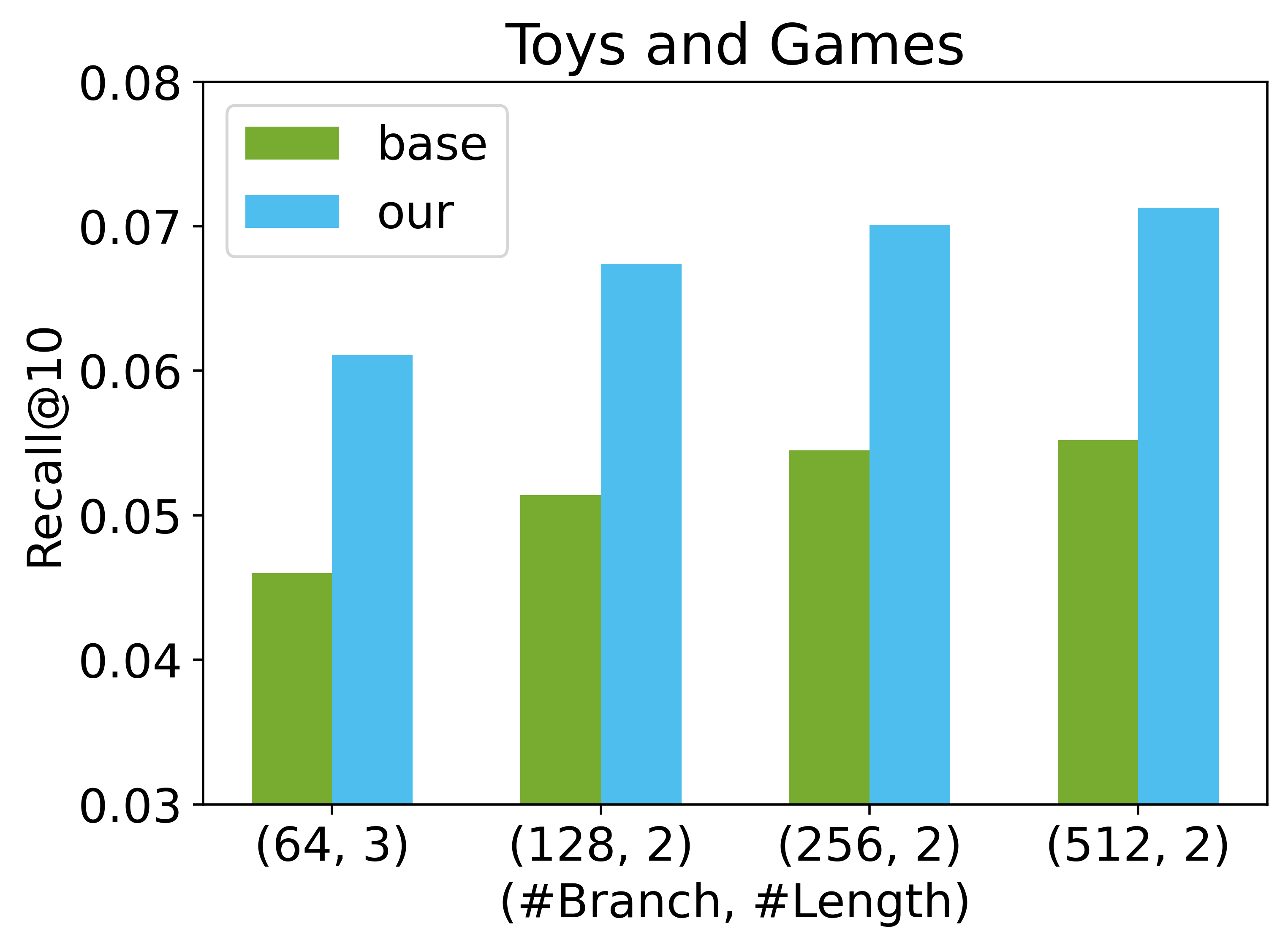}
        \caption{Toys and Games}
        \label{fig:cluster_toy}
	\end{subfigure}

    \caption{Impact of branch number $k$, ranging from 64 to 512, in terms of Recall@10. The corresponding identifier length $l$ is also annotated.
}
    \label{fig:cluster}
    \end{center}
    %\vspace{-0.5cm}
\end{figure}

\vpara{Cluster Number.} We investigated the impact of employing varying branch numbers $k$ on model performance. The increase in branch number $k$ leads to a corresponding decrease in the length $l$ of the item identifier according to the total number of items.
The experiment is conducted on two datasets, Beauty and Toys.
aThe results are illustrated in Fig.~\ref{fig:cluster}.
We observe that as $k$ increases from 64 to 512, the model performance of the base and ours both monotonically increase on the Toys dataset. However, an interesting trend emerged on the Beauty dataset, where we observed a decline in performance as $k$ increased from 256 to 512. The performance improvement resulting from increasing $k$ can be attributed to the reduction in identifier length $l$, but an excessively larger $k$ can lead to a decline in model performance and increase inference time. Moreover, our methods always show a better performance than the base one, which suggests the superiority of two-decoder architecture.

% \subsection{Qualitative Analysis (RQ3)}

\section{Conclusion and Future Work}
In this paper, we introduce a novel framework, EAGER, designed to integrate behavioral and semantic information for unified generative recommendation. EAGER comprises three key components: (1) a two-stream generation architecture that combines behavioral and semantic information to enhance item recommendation, (2) a global contrastive task with a summary token to capture global knowledge for improved auto-regressive generation quality, and (3) a semantic-guided transfer task that facilitates interactions across two decoders and  their features. Extensive comparisons with state-of-the-art methods and detailed analyses demonstrate the effectiveness and robustness of EAGER. In future work, we plan to further enhance generative recommendation models by incorporating large language models and multimodal AI techniques~\cite{TenChallenges}.

\begin{acks}
We thank MindSpore\footnote{https://www.mindspore.cn} for the partial support of this work, which is a new deep learning computing framework.
\end{acks}